\documentclass[12pt]{article}
\usepackage{amsmath}
\input{epsf}
\def\ltap{\raisebox{-.4ex}{\rlap{$\,\sim\,$}} \raisebox{.4ex}{$\,<\,$}}
\def\gtap{\raisebox{-.4ex}{\rlap{$\,\sim\,$}} \raisebox{.4ex}{$\,>\,$}}

\newcommand\as{\alpha_{\mathrm{S}}}

\def\beeq{\begin{eqnarray}}
\def\eeeq{\end{eqnarray}}

\def\to{\rightarrow}
\def\nn{\nonumber}

\def\qt{q_{\perp}}
\def\res{{\rm res.}}
\def\ms{${\overline {\rm MS}}$}
\def\msbar{{\overline {\rm MS}}}

\begin{document}
\begin{titlepage}
\renewcommand{\thefootnote}{\fnsymbol{footnote}}
\begin{flushright}
    CERN--TH/2000-244\\ hep-ph/0008184
     \end{flushright}
\par \vspace{10mm}
\begin{center}
{\Large \bf
Universality of non-leading logarithmic contributions
\\[1.ex]
in transverse-momentum distributions}
\footnote{This work was supported in part 
by the EU Fourth Framework Programme ``Training and Mobility of Researchers'', 
Network ``Quantum Chromodynamics and the Deep Structure of
Elementary Particles'', contract FMRX--CT98--0194 (DG 12 -- MIHT).}
\end{center}
\par \vspace{2mm}
\begin{center}
{\bf Stefano Catani${}^{(a)}$~\footnote{On leave of absence 
from INFN, Sezione di Firenze, Florence, Italy.}, Daniel de Florian${}^{(b)}$}
\hskip .2cm
and
\hskip .2cm
{\bf Massimiliano Grazzini${}^{(b)}$}~\footnote{Work supported by the Swiss
  National Foundation.}\\

\vspace{5mm}

${}^{(a)}$Theory Division, CERN, CH 1211 Geneva 23, Switzerland \\

${}^{(b)}$Institute for Theoretical Physics, ETH-H\"onggerberg, 
CH 8093 Z\"urich, Switzerland

\vspace{5mm}

\end{center}

\par \vspace{2mm}
\begin{center} {\large \bf Abstract} \end{center}
\begin{quote}
\pretolerance 10000

We consider the resummation of the logarithmic contributions to the 
region of small transverse momenta in the distributions
of  high-mass systems (lepton pairs, vector bosons, Higgs particles, $\dots$)
produced in hadron collisions. 
We point out that the resummation formulae that are usually
used to compute the distributions in perturbative QCD involve
process-dependent form factors and coefficient functions.
We present a new universal form of the resummed distribution, in which the 
dependence on the process is embodied in a single perturbative factor.
The new form simplifies the calculation of non-leading logarithms at 
higher perturbative orders. It can also be useful
to systematically implement process-independent non-perturbative effects
in transverse-momentum distributions.
We also comment on the dependence of these distributions
on the factorization and renormalization scales.

\end{quote}

\vspace*{\fill}
\begin{flushleft}
     CERN--TH/2000-244 \\August 2000 

\end{flushleft}
\end{titlepage}

\setcounter{footnote}{1}

\renewcommand{\thefootnote}{\fnsymbol{footnote}}

The properties of the transverse-momentum distributions of systems of high mass
$Q$ produced at high-energy hadron colliders are important for QCD studies
and for physics studies beyond the Standard Model
(see, e.g., Refs.~\cite{Catani:2000jh}--\cite{Baur:2000xd}).
On the theoretical side the computation of these distributions is complicated 
by the presence of large logarithmic contributions of the type
$\ln Q^2/\qt^2$, which spoil the convergence of fixed-order perturbative
calculations in the region of small transverse momenta $\qt$. The
logarithmically-enhanced terms have to be evaluated at higher
perturbative orders, and possibly resummed to all orders in the QCD coupling
$\as$. The all-order resummation formalism was developed in the eighties
\cite{Dokshitzer:1980hw}--\cite{Davies:1984hs}.
The structure of the resummed distribution is given in terms of a 
transverse-momentum form factor and of process-dependent contributions.

The transverse-momentum form factor is sometimes supposed to be universal, 
that is, independent of the process. Performing a process-independent 
calculation at relative order $\as^2$, two of us have recently
shown~\cite{defgraz} in an explicit way that the form factor depends on the
process.  In this note, we give a general interpretation of the results in
Ref.~\cite{defgraz}. We point out that this process-dependent features
persist to higher orders, at least in the final form in which 
the resummed transverse-momentum distribution is {\em usually} organized.
We also present a new form of the resummed distribution, where all the
logarithmic contributions are embodied in universal (process-independent)
factors. 

We consider the inclusive hard-scattering process
\begin{equation}
\label{process}
h_1(p_1) + h_2(p_2) \to F(Q^2,\qt^2) + X \;\;,
\end{equation}
where the triggered final-state system $F$ is produced by the collision of the
two hadrons $h_1$ and $h_2$ with momenta $p_1$ and  $p_2$, respectively.
We denote by $\sqrt s$ the centre-of-mass energy of the colliding hadrons
$(s= (p_1+p_2)^2 \simeq 2p_1p_2)$.
The final state $F$ is a generic system of non-strongly interacting particles
\footnote{We do not consider the production of strongly interacting particles
(hadrons, jets, heavy quarks, ...), because in this case the resummation
formalism of small-$\qt$ logarithms has not yet been fully developed.},
such as {\em one} or {\em more} vector bosons $(\gamma^*, W, Z, \dots)$, 
Higgs particles ($H$) and so forth. 

To simplify the discussion we limit ourselves to the case in which only the
total invariant mass $Q$ and transverse momentum $\qt$ (with respect to the
direction of the colliding hadrons) of the system $F$ are measured. The
extension to more general kinematic configurations, in which, for instance,
the rapidities are measured, is straightforward (see the discussion
below Eq.~(\ref{b3dep})).
We also assume that at the parton level the system $F$ is produced with
vanishing $\qt$ (i.e. with no accompanying final-state radiation)
in the leading-order (LO) approximation, so that the
corresponding cross section is $d\sigma^{(LO)}/d\qt^2 \propto \delta(\qt^2)$. 
Since $F$ is colourless, the LO partonic subprocess is either
$q{\bar q}$ {\em annihilation}, as in the case of $\gamma^*, W$ and $Z$ 
production, or $gg$ {\em fusion}, as in the case of the production of a Higgs
boson $H$.

The transverse-momentum cross section for the process
in Eq.~(\ref{process}) can be written as~\cite{Parisi:1979se, Collins:1981uk,
Kodaira:1982nh}
\begin{equation}
\label{Fdec}
\frac{d\sigma_{F}}{dQ^2 \;d\qt^2} = 
\left[ \frac{d\sigma_{F}}{dQ^2 \;d\qt^2} \right]_{\res}
+ \left[ \frac{d\sigma_{F}}{dQ^2 \;d\qt^2} \right]_{{\rm fin.}} \;\;.
\end{equation}
Both terms on the right-hand side are obtained as convolutions of
partonic cross sections and the parton distributions $f_{a/h}(x,Q^2)$ 
($a=q_f, {\bar q}_f, g$ is the parton label) of the
colliding hadrons\footnote{Throughout the paper we always use
parton densities as defined in the \ms\ factorization scheme and $\as(q^2)$ is
the QCD running coupling in the \ms\ renormalization scheme.}. 
The distinction between the two terms is purely theoretical.
The partonic cross section that enters in the resummed part (the first term
on the right-hand side) contains all the logarithmically-enhanced
contributions $\as^n \ln^m Q^2/\qt^2$. Thus, this part has to be evaluated by 
resumming the logarithmic terms to all orders in perturbation theory. On the
contrary, the partonic cross section in the second term on the right-hand side
is finite order-by-order in perturbation theory when $\qt \to 0$. It can thus
be computed by truncating the perturbative expansion at a given fixed order in
$\as$. 

The finite component of the transverse-momentum cross section
is obviously process-dependent, and we have nothing to add on it in this paper.
In the following we discuss the structure of the resummed part.

The resummed component is\footnote{As discussed at the end of the paper,
this expression can be generalized
to include the dependence on the renormalization and factorization scales 
$\mu_R$ and $\mu_F$.}
\beeq
\label{resgen}
\left[ \frac{d\sigma_{F}}{dQ^2 \;d\qt^2} \right]_{\res} &=& \sum_{a,b}
\int_0^1 dx_1 \int_0^1 dx_2 \int_0^\infty db \; \frac{b}{2}\; J_0(b \qt) 
\;f_{a/h_1}(x_1,b_0^2/b^2) \; f_{b/h_2}(x_2,b_0^2/b^2) \nn \\
&\cdot& W_{ab}^{F}(x_1 x_2 s; Q, b) \;\;.
\eeeq
The Bessel function $J_0(b \qt)$ and the coefficient $b_0=2e^{-\gamma_E}$
($\gamma_E=0.5772\dots$ is the Euler number) have a kinematical origin.
To correctly take into account the kinematics constraint of transverse-momentum
conservation, the resummation procedure has to be carried out in the 
impact-parameter space. The transverse-momentum cross section (\ref{resgen})
is then obtained by performing the inverse Fourier (Bessel) transformation 
with respect to the impact parameter $b$.
The factor $W_{ab}^{F}$ is the perturbative and 
process-dependent partonic cross section that embodies the all-order 
resummation of the large logarithms $\ln Q^2b^2$ (the limit $\qt \ll Q$ 
corresponds to $Qb \gg 1$, because $b$ is the variable conjugate to $\qt$).

The resummed partonic cross section is
{\em usually} (see, e.g., the list of references in 
Sections~5.1 and 5.3 of Ref.~\cite{Catani:2000jh}) written in the following 
form:
\beeq
\label{nonunw}
W_{ab}^{F}(s; Q, b) &=& \sum_c \int_0^1 dz_1 \int_0^1 dz_2 
\; C_{ca}^{F}(\as(b_0^2/b^2), z_1) \; C_{{\bar c}b}^{F}(\as(b_0^2/b^2), z_2)
\; \delta(Q^2 - z_1 z_2 s) \nn \\
&\cdot& \sigma_{c{\bar c}}^{(LO) \,F}(Q^2) \;S_c^{F}(Q,b) \;\;.
\eeeq
Here, $\sigma_{c{\bar c}}^{(LO) \,F}$ is the cross section
(integrated over $\qt$) for the LO partonic subprocess
$c + {\bar c} \to F$, where $c,{\bar c}=q,{\bar q}$ (the quark $q_f$ and
the antiquark ${\bar q}_{f'}$ can possibly have different flavours $f,f'$)
or $c,{\bar c}=g,g$. The expression $\sigma_{c{\bar c}}^{(LO) \,F}$ can 
include an overall factor $\as^p(Q^2)$, as in the case of $g+g \to H$ 
through a triangular quark loop where $p=2$.  
The term $S_c^{F}(Q,b)$ is the quark $(c=q)$ or
gluon $(c=g)$ Sudakov form factor. The resummation of the logarithmic
contributions is achieved by exponentiation
\cite{Dokshitzer:1980hw}--\cite{Bassetto:1980nt}, 
that is by showing \cite{Collins:1981uk, Kodaira:1982nh} that the form factor 
can be expressed as 
\begin{equation}
\label{formfact}
S_c(Q,b) = \exp \left\{ - \int_{b_0^2/b^2}^{Q^2} \frac{dq^2}{q^2} 
\left[ A_c(\as(q^2)) \;\ln \frac{Q^2}{q^2} + B_c(\as(q^2)) \right] \right\} 
\;\;, 
\end{equation}
with $c=q$ or $g$. The functions $A_c(\as), B_c(\as)$, as well as the 
coefficient functions $C_{ab}(\as, z)$ in Eq.~(\ref{nonunw}), contain no
$\ln Q^2b^2$ terms and are perturbatively computable according to the
power expansions\footnote{Note that in Refs.~\cite{Davies:1984hs, defgraz}
the perturbative coefficients are normalized to powers of $\as/2\pi$ rather 
than $\as/\pi$.}
\beeq
\label{aexp}
A_c(\as) &=& \sum_{n=1}^\infty \left( \frac{\as}{\pi} \right)^n A_c^{(n)} 
\;\;, \\
\label{bexp}
B_c(\as) &= &\sum_{n=1}^\infty \left( \frac{\as}{\pi} \right)^n B_c^{(n)}
\;\;, \\
\label{cexp}
C_{ab}(\as,z) &=& \delta_{ab} \,\delta(1-z) + 
\sum_{n=1}^\infty \left( \frac{\as}{\pi} \right)^n C_{ab}^{(n)}(z) \;\;.
\eeeq
The knowledge of the coefficients $A^{(1)}$ leads to the resummation of
the leading logarithmic (LL) contributions. Analogously, the coefficients 
$\{ A^{(2)}, B^{(1)}, C^{(1)} \}$ give the next-to-leading logarithmic (NLL)
terms, the coefficients $\{ A^{(3)}, B^{(2)}, C^{(2)} \}$ give the 
next-to-next-to-leading logarithmic (NNLL) terms, and so forth. The 
coefficients $A^{(1)}, A^{(2)}, B^{(1)}$ are known both for
the quark \cite{Kodaira:1982nh} and for the gluon \cite{Catani:1988vd} 
form factors
\beeq
&&A_q^{(1)} = C_F \;\;, \quad \quad \;\;A_g^{(1)} = C_A \;, \nn\\
\label{ffcoef}
&&A_q^{(2)} = \frac{1}{2} C_F K \;\;, \quad A_g^{(2)} = \frac{1}{2} C_A K \;,\\
&&B_q^{(1)} = - \frac{3}{2} C_F \;\;, \quad 
B_g^{(1)} = - \frac{1}{6} (11 C_A - 2 N_f) \;, \nn
\eeeq
where 
\begin{equation}
K = \left( \frac{67}{18} - \frac{\pi^2}{6} \right) C_A - \frac{5}{9} N_f \;\;.
\end{equation}

The best studied example among the processes in Eq.~(\ref{process})
is lepton-pair Drell--Yan (DY) 
production through the LO partonic subprocess $q + {\bar q} \to \gamma^* 
({\rm or} \;W,Z) \to l + l'$. In this case the NNLL coefficient $B^{(2)}$
was computed by Davies and Stirling \cite{Davies:1984hs}:
\begin{equation}
\label{b2dy}
B_q^{(2) \,DY} = 
C_F^2\left(\frac{\pi^2}{4}-\frac{3}{16}-3\zeta_3\right)
+C_F\,C_A\left(\frac{11}{36}\pi^2-\frac{193}{48}+\frac{3}{2}\zeta_3\right)
+ C_F\,N_f\left(\frac{17}{24}-\frac{\pi^2}{18}\right)\,,
\end{equation}
where $\zeta_n$ is the Riemann $\zeta$-function $(\zeta_3=1.202\dots)$.
The coefficient $B^{(2)}$ for Higgs boson production has recently been computed
by two of us~\cite{defgraz}. The corresponding LO partonic subprocess is gluon
fusion, $g + g \to H$, through a massive-quark loop. In the limit of infinite
quark mass the value of $B^{(2)}$ is~\cite{defgraz} 
\begin{equation}
\label{b2h}
B_g^{(2) \,H}=C_A^2\left(\frac{23}{24}+
\frac{11}{18}\pi^2-\frac{3}{2}\zeta_3\right)
+\frac{1}{2} C_F\,N_f-C_A\,N_f\left(\frac{1}{12}+\frac{\pi^2}{9} \right)
-\frac{11}{8} C_F C_A\, .
\end{equation}

The main issue that we want to discuss in this paper regards the
process dependence of the various factors in the resummation formula 
(\ref{nonunw}).
As denoted by the superscripts $F$ in the
various terms on the right-hand side, the coefficient functions $C_{ab}^F$ 
depend on the process. This is confirmed by the calculations of the
coefficients $C_{ab}^{(1) \,F}$, performed in the literature for several
processes
\footnote{A general expression for the coefficients $C_{ab}^{(1) \,F}$ 
in terms of the one-loop matrix element of the corresponding process is
given in Eq.~(17) of Ref.~\cite{defgraz}.}
\cite{Davies:1984hs} \cite{Altarelli:1984pt}--\cite{Balazs:1999bm}.
The form factor $S_c(Q,b)$ that enters 
Eq.~(\ref{nonunw})
is (often) {\em supposed} to be universal (this is the reason why it is named
quark or gluon form factor rather than DY, $\gamma \gamma$, $WZ$, $H, \dots$
form factor). However, this is not the case: the form factor $S_c^F(Q,b)$
in Eq.~(\ref{nonunw}) is {\em process-dependent}.
In the following, we first present a {\em universal} (process-independent)
version of the resummation formula (\ref{nonunw}) and we sketch its physical 
origin. We then clarify the relation between Eq.~(\ref{nonunw}) and our 
process-independent version.

The process-independent resummation formula is
\beeq
\label{unw}
W_{ab}^{F}(s; Q, b) &=& \sum_c \int_0^1 dz_1 \int_0^1 dz_2 
\; C_{ca}(\as(b_0^2/b^2), z_1) \; C_{{\bar c}b}(\as(b_0^2/b^2), z_2)
\; \delta(Q^2 - z_1 z_2 s) \nn \\
&\cdot& \sigma_{c{\bar c}}^{F}(Q^2, \as(Q^2)) \;S_c(Q,b) \;\;.
\eeeq
It formally differs from Eq.~(\ref{nonunw}) by the replacement
$\sigma_{c{\bar c}}^{(LO) \,F}(Q^2) \to \sigma_{c{\bar c}}^{F}(Q^2, \as(Q^2))$.
While $\sigma_{c{\bar c}}^{(LO) \,F}(Q^2)$ is the cross section for the
LO partonic subprocess, $\sigma_{c{\bar c}}^{F}(Q^2, \as(Q^2))$ includes
higher-order QCD corrections to it, according to
\begin{equation}
\sigma_{c{\bar c}}^{F}(Q^2, \as(Q^2)) = \sigma_{c{\bar c}}^{(LO) \,F}(Q^2) \;
H_c^{F}(\as(Q^2)) \;\;,
\end{equation}
where the function $H_c^{F}(\as)$ has a perturbative expansion
similar to Eqs.~(\ref{aexp})--(\ref{cexp}):
\begin{equation}
\label{hexp}
H_c^{F}(\as) = 1 +  
\sum_{n=1}^\infty \left( \frac{\as}{\pi} \right)^n H_c^{(n) \,F} \;\;.
\end{equation}
Note that the function $H_c^{F}(\as)$ depends on the process.
Nonetheless, its introduction is sufficient to transform the
process-dependent form factor $S_c^F$ and coefficient functions $C_{ca}^F$
of Eq.~(\ref{nonunw}) into the process-independent 
form factor $S_c$ {\em and} coefficient functions $C_{ca}$
of Eq.~(\ref{unw}).

\begin{figure}[htb]
\begin{center}
\begin{tabular}{c}
\epsfxsize=10truecm
\epsffile{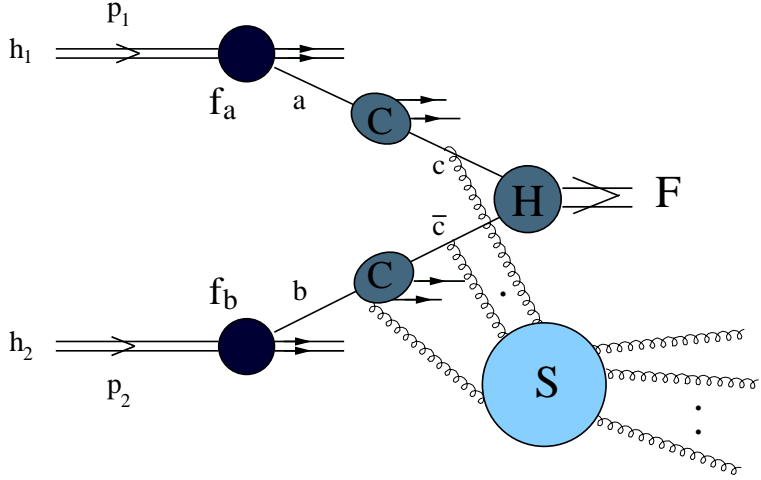}\\
\end{tabular}
\end{center}
\caption{\label{fig:smallqt}{\em Diagrammatic structure of the various factors
that enter the process-independent resummation formula (\ref{unw}).}}
\end{figure}

The resummation formula in Eq.~(\ref{unw}), which can be derived by the
customary resummation methods 
\cite{Dokshitzer:1980hw}--\cite{Kodaira:1982nh} \cite{Catani:1988vd},
has a simple physical origin.
When the final-state system $F$ is kinematically
constrained to have a small transverse momentum, 
the emission of accompanying radiation is strongly inhibited, so that only
soft and collinear partons (i.e. partons with low
transverse momenta $q_t$) can be radiated in the final state 
(Fig.~\ref{fig:smallqt}). The
process-dependent factor $H_c^{F}(\as(Q^2))$ embodies hard contributions
produced by virtual corrections at transverse-momentum
scales $q_t \sim Q$. The form factor $S_c(Q,b)$ contains 
real and virtual contributions due to soft (the function $A_c(\as)$ in
Eq.~(\ref{formfact})) and flavour-conserving collinear (the function
$B_c(\as)$ in Eq.~(\ref{formfact})) radiation at scales 
$Q \gtap q_t \gtap 1/b$. At very low momentum scales, $q_t \ltap 1/b$,
real and virtual soft-gluon corrections cancel because the cross section is 
infrared safe, and only real and virtual contributions due to collinear 
radiation remain (the coefficient functions $C_{ab}(\as(b_0^2/b^2),z)$).
Note that $S_c(Q,b)$ and $C_{ab}(\as(b_0^2/b^2),z)$ are process-independent
and only depend on the flavour and colour
charges of the QCD partons.

The two versions (\ref{nonunw}) and (\ref{unw}) of the resummation formula
can formally be related as follows. We use the renormalization-group identity
\begin{equation}
\label{rgid}
g_1(\as(Q^2)) = 
\exp \left\{ \int_{b_0^2/b^2}^{Q^2} \frac{dq^2}{q^2} \;g_2(\as(q^2)) \right\}
\; g_1(\as(b_0^2/b^2)) \;\;,
\end{equation}
which is valid when
\begin{equation}
g_2(\as) = \beta(\as) \frac{d\ln g_1(\as)}{d\ln \as} \;\;,
\end{equation}
where $\beta(\as)$ is the QCD $\beta$-function
\begin{equation}
\frac{d\ln \as(q^2)}{d\ln q^2} = \beta(\as(q^2)) \;\;, 
\end{equation}
\begin{equation}
\beta(\as) = - \beta_0 \frac{\as}{\pi} - \beta_1 \left(\frac{\as}{\pi}\right)^2
+ \dots \;\;, \quad
12 \beta_0 = 11 C_A - 2 N_f \;\;.
\end{equation}
Then, setting $g_1(\as(Q^2))=H_c^{F}(\as(Q^2))$ and inserting the right-hand
side of Eq.~(\ref{rgid}) in Eq.~(\ref{unw}), we immediately obtain 
Eq.~(\ref{nonunw}). More precisely, the process-independent resummation
formula in Eq.~(\ref{unw}) implies the customary version in
Eq.~(\ref{nonunw}), provided the perturbative function $B_c(\as)$ 
in the form factor $S_c^F$ (see Eq.~(\ref{formfact})) and the coefficient
functions $C_{ab}^F$ are related to their process-independent
analogues by the following all-order relations
\beeq
\label{cdep}
C_{ab}^F(\as, z) &=& \left[ H_a^F(\as) \right]^{1/2} \;C_{ab}(\as, z) \;\;, \\
\label{bdep}
B_c^F(\as) &=& B_c(\as) - \beta(\as) \;\frac{d\ln H_c^F(\as)}{d\ln \as} \;\;.
\eeeq
While the perturbative function $A_c(\as)$ and the first-order coefficient
$B_c^{(1)}$ of the function $B_c(\as)$ are process-independent, the result
in Eqs.~(\ref{bdep}) and (\ref{cdep}) shows that the coefficients
$B_c^{(2)}, B_c^{(3)}, \dots$ and $C_{ab}^{(1)}(z), C_{ab}^{(2)}(z), \dots$
in Eq.~(\ref{nonunw}) do depend on the process. The process dependence of the
first few coefficients is explicitly given by
\beeq
\label{c1dep}
C_{ab}^{(1) \,F}(z) &=& C_{ab}^{(1)}(z) + \delta_{ab} \;\delta(1-z) \;
\frac{1}{2} \; H_a^{(1) \,F} \;\;, \\
\label{c2dep}
C_{ab}^{(2) \,F}(z) &=& C_{ab}^{(2)}(z) + \frac{1}{2} \; H_a^{(1) \,F}
\;C_{ab}^{(1)}(z) + \delta_{ab} \;\delta(1-z) \;\frac{1}{2}
\;\left[
H_a^{(2) \,F} - \frac{1}{4} \left( H_a^{(1) \,F} \right)^2 \right] \;, \\
 \label{b2dep}
B_c^{(2) \,F} &=&B_c^{(2)} + \beta_0 H_c^{(1) \,F} \;\;, \\
\label{b3dep}
B_c^{(3) \,F} &=& B_c^{(3)} + \beta_1 H_c^{(1) \,F} + 
2 \beta_0 \left[
H_c^{(2) \,F} - \frac{1}{2} \left( H_c^{(1) \,F} \right)^2 \right] \;\;. 
\eeeq

Note that the process dependence of the resummation formula (\ref{nonunw})
is not simply embodied in pure numerical coefficients. For example,
considering the Higgs production mechanism $g+g \to H$ through a massive-quark
loop, the corresponding coefficient $H_g^{(1) \,H}$ and, hence, the form-factor
coefficient $B_g^{(2) \,H}$ depend on the mass of the quark in the triangular
loop. Moreover, the results discussed so far can straightforwardly be
extended to more general configurations, in which several kinematical
variables (and not only the invariant mass $Q$) of the final-state system
$F$ are measured. For example, when $F$ is a pair of vector bosons, we can
consider the corresponding transverse-momentum cross section at fixed rapidities
of the vector bosons. The extension simply amounts to including the
dependence on the kinematics of the system $F$ in the cross section factor
$\sigma_{c{\bar c}}^{F}$ (and, hence, in $\sigma_{c{\bar c}}^{(LO) \,F}$ and
$H_c^{F}$) of the process-independent resummation formula (\ref{unw}). In this
general case, Eqs.~(\ref{bdep}) and (\ref{cdep}) imply that the form
factor $S_c^F$ and the coefficient functions $C_{ab}^F$ in Eq.~(\ref{nonunw})
depend on the kinematics of the system $F$ in a non-trivial manner.

The process-independent resummation formula (\ref{unw}) can alternatively
be used to relate the factors on the right-hand side of Eq.~(\ref{nonunw})
for different processes. Considering Eqs.~(\ref{bdep}) and (\ref{cdep})
for two different processes $F$ and $\widetilde F$, we obtain
\beeq
\label{cff}
C_{cb}^F(\as, z) &=& \left[ H_c^{F{\widetilde F}}(\as) 
\right]^{1/2} \;C_{cb}^{\widetilde F}(\as, z) \;\;,\\
\label{bff}
B_c^F(\as) &=& B_c^{\widetilde F}(\as) - \beta(\as) 
\;\frac{d\ln H_c^{F{\widetilde F}}(\as)}{d\ln \as} \;\;, 
\eeeq
where $H^{F{\widetilde F}}=H^F/H^{\widetilde F}$. The perturbative expansion 
of these equations obviously leads to relations that are similar to those
in Eqs.~(\ref{c1dep})--(\ref{b2dep}):
\beeq
\label{c1ff}
C_{cb}^{(1) \,F}(z) &\!\!\!=\!\!\!& C_{cb}^{(1) \,{\widetilde F}}(z) 
+ \delta_{cb} \;\delta(1-z) \;
\frac{1}{2} \, H_c^{(1) \,F{\widetilde F}} \;\;, \\
\label{c2ff}
C_{cb}^{(2) \,F}(z) &\!\!\!=\!\!\!& C_{cb}^{(2) \,{\widetilde F}}(z) + 
\frac{1}{2} H_c^{(1) \,F{\widetilde F}}
C_{cb}^{(1) \,{\widetilde F}}(z) + \delta_{cb} \,\delta(1-z) \frac{1}{2}
\left[
H_c^{(2) \,F{\widetilde F}} - 
\frac{1}{4} \left( H_c^{(1) \,F{\widetilde F}} \right)^2 \right] , \\
 \label{b2ff}
B_c^{(2) \,F} &\!\!\!=\!\!\!&B_c^{(2) \,{\widetilde F}} + 
\beta_0 H_c^{(1) \,F{\widetilde F}} \;\;, \\
\label{b3ff}
B_c^{(3) \,F} &\!\!\!=\!\!\!& B_c^{(3) \,{\widetilde F}} + 
\beta_1 H_c^{(1) \,F{\widetilde F}} + 
2 \beta_0 \left[
H_c^{(2) \,F{\widetilde F}} 
- \frac{1}{2} \left( H_c^{(1) \,F{\widetilde F}} \right)^2 \right] \;\;. 
\eeeq
Since the function $H^{F{\widetilde F}}$ does not appear in 
Eq.~(\ref{nonunw}), within the framework of the process-dependent resummation
formulae, Eqs.~(\ref{cff})--(\ref{b3ff}) have to be regarded as 
non-linear relations between the form factors and coefficient functions of 
different processes. For instance, having computed the first-order
coefficient $C_{ab}^{(1)}$ for two processes, one can check Eq.~(\ref{c1ff}) 
and extract $H_c^{(1) \,F{\widetilde F}}$. 
Then, the second-order coefficients $B_c^{(2)}$ of these
processes must be related by Eq.~(\ref{b2ff}). This constraint can either be
checked or used to compute $B_c^{(2) \,{\widetilde F}}$ from 
$B_c^{(2) \,F}$.

The relation in Eq.~(\ref{c1ff}) (or Eq.~(\ref{c1dep})) can straightforwardly
be checked by 
comparing the known first-order coefficients $C_{ab}^{(1) \,F}(z)$ for
the production of DY lepton pairs \cite{Davies:1984hs, Altarelli:1984pt,
Davies:1985sp, Balazs:1997xd}, diphotons \cite{Balazs:1998xd} and $ZZ$ pairs
\cite{Balazs:1999bm}. In particular, we have
\begin{equation}
\label{h1diff}
H_q^{(1) \,F \,DY} = H_q^{(1) \,F} - H_q^{(1) \,DY} 
= \frac{1}{2} \;C_F \left( {\cal V}^{F} - \pi^2 + 8
\right) \;\;,
\quad F=\gamma\gamma ,ZZ
 \;\;,
\end{equation}
where ${\cal V}^{\gamma\gamma}$ and ${\cal V}^{ZZ}$ are given in Eq.~(11)
of Ref.~\cite{Balazs:1998xd} and in Eq.~(8)\footnote{As pointed out
in Ref.~\cite{defgraz}, the actual expression in Eq.~(8) of 
Ref.~\cite{Balazs:1999bm} is not correct.}
of Ref.~\cite{Balazs:1999bm}, respectively.
The relation (\ref{b2ff}) (or (\ref{b2dep})) for the NNLL coefficient
$B^{(2) \,F}$ has explicitly been derived by the process-independent 
calculation at ${\cal O}(\as^2)$ performed in Ref.~\cite{defgraz}. 

The reasoning used to obtain Eq.~(\ref{nonunw}) from Eq.~(\ref{unw})
can also be used to show that Eq.~(\ref{unw}) is invariant under the 
transformation
\beeq
H_c^{F}(\as(Q^2)) & \to & H_c^{F}(\as(Q^2)) \; \left[ g(\as(Q^2)) \right]^{-1} 
\;, \nn \\
S_c(Q,b) & \to & S_c(Q,b) \;
\exp \left\{ \int_{b_0^2/b^2}^{Q^2} \frac{dq^2}{q^2} 
\;\beta(\as(q^2)) \;\frac{d\ln g(\as(q^2))}{d\ln \as(q^2)} \right\} \;,\\
C_{ab}(\as(b_0^2/b^2),z) & \to & C_{ab}(\as(b_0^2/b^2),z) \;
\left[ g(\as(b_0^2/b^2)) \right]^{1/2} \;, \nn 
\eeeq
where $g(\as) = 1 + {\cal O}(\as)$ is an arbitrary perturbative function.
This renormalization-group symmetry of the resummation formula (\ref{unw})
implies that its factors, $H_c^{F}, S_c$ (more precisely, the function $B_c$) 
and $C_{ab}$, although they are
process-independent, are not unambiguously
computable (defined) order by order in perturbation theory.
The same conclusion can be reached by expanding Eq.~(\ref{unw}) in powers
of $\as$. At any given order in $\as$, the coefficients of the various 
powers of $\ln Q^2b^2$ depend on the unknowns $H_c^{(n) \,F}, C_{ab}^{(n)}, 
B_c^{(n)}$, but the number of such coefficients is less that the number of
unknowns. For instance, at the first relative order in $\as$, the coefficient
of $\ln^2 Q^2b^2$ determines $A^{(1)}$, the coefficient of $\ln Q^2b^2$ 
determines $B^{(1)}$, while the coefficient of the constant term is a linear
combination of $H^{(1) \,F}$ and $C_{ab}^{(1)}$. 

\setcounter{footnote}{0}

The perturbative ambiguity of the decomposition of the right-hand side of
Eq.~(\ref{unw}) in the factors $H_c^{F}, S_c$ and $C_{ab}$ is a consequence of 
the fact that the transverse-momentum cross section is not a collinear-safe 
quantity. The effect of collinear radiation at low transverse-momentum scales,
$q_t \ltap 1/b$, is divergent in perturbation theory. The (arbitrary)
regularization procedure of these divergences introduces some ambiguity 
in the definition of the coefficient functions $C_{ab}$. Then, the ambiguity
propagates to $S_c$ and $H_c^{F}$ through collinear evolution (see
Eq.~(\ref{rgid})). In this way of thinking, the ambiguity is similar to that 
encountered in the definition of the parton densities. 
As the parton densities have to be defined by fixing a factorization scheme
(e.g. the \ms\ scheme or the DIS scheme), the factors on the
right-hand side of Eq.~(\ref{unw}) has to be defined by choosing a `resummation
scheme'. Note that the choice of a `resummation scheme' amounts to
defining $H_c^{F}$ (or $C_{ab}$) for a {\em single} process\footnote{More 
precisely,
$H_c^{F}$ has to be defined for two processes: one process that 
is controlled, at LO, by $q{\bar q}$ annihilation and another process that  
is controlled, at LO, by $gg$ fusion.}. Having done that, the process-dependent
factor $H_c^{F}$ and the universal factors $S_c$ and $C_{ab}$ in Eq.~(\ref{unw})
are unambiguously determined for any other process in Eq.~(\ref{process}).

We can suggest three examples of `resummation schemes', that is three 
prescriptions to use the process-independent factorization formula (\ref{unw})
in practice. A first `short-cut' (because the DY transverse-momentum
distribution is best studied)
prescription, 
which we name `DY resummation scheme', is obtained by setting 
$H_q^{DY}(\as)\equiv1$ in the case of the DY cross section (integrated over
the rapidity of the vector boson). This unambiguously fixes the 
process-independent form factor $S_c$ and coefficients functions $C_{ab}$
as those determined from the DY process. In particular, the coefficient
$B_q^{(2)}$ of the quark form factor is that in Eq.~(\ref{b2dy}), and the
quark coefficient functions $C_{qb}^{(1)}(z)$ $(b=q,{\bar q},g)$
are those computed in
Refs.~\cite{Davies:1984hs, Altarelli:1984pt, Davies:1985sp, Balazs:1997xd}
for the DY process.
Any other $q{\bar q}$-initiated cross section is then obtained by using
Eq.~(\ref{unw}) with a perturbatively computable hard function $H_q^{F}(\as)$.
This prescription can be extended to $gg$-initiated cross sections
by setting $H_g^{H}(\as)\equiv1$ in the case of Higgs production via $gg$ fusion.
Thus, in the case of an infinitely-massive quark in the loop (the definition 
of the scheme will be different by keeping the mass of the quark finite), the
coefficient $B_g^{(2)}$ of the gluon form factor is that given in 
Eq.~(\ref{b2h}) and the gluon coefficient functions $C_{gb}^{(1)}(z)$ 
$(b=q,{\bar q},g)$ are those computed in Ref.~\cite{Kauffman:1992cx}.

A second prescription, which is in some sense more physical, exploits the
fact that the first moment (with respect to $Q^2/s$) of the
flavour non-singlet (NS) hadronic (and partonic) cross sections defines
collinear-safe quantities.
This property follows from fermion-number conservation\footnote{In the case of
the Altarelli--Parisi evolution of the parton densities, 
fermion-number conservation implies the vanishing of the first moment of the
Altarelli--Parisi probabilities.}. Thus, we can define a `NS resummation
scheme' by fixing the overall normalization of the NS coefficient function
$C_{ab}(\as,z)$ in Eq.~(\ref{unw}) in such a way that its first moment
(with respect to $z$) vanishes. In this scheme (the subscript NS labels the
scheme choice)
\begin{equation}
\int_0^1 dz \; \left[ \; C_{qq,NS}(\as,z) - C_{q{\bar q},NS}(\as,z) \;\right] 
\equiv 1 \;\;,
\end{equation}
and thus, using the DY coefficients functions from 
Refs.~\cite{Davies:1984hs, Altarelli:1984pt, Davies:1985sp, Balazs:1997xd}
and Eq.~(\ref{c1dep}), we obtain
\begin{equation}
\label{h1dyns}
H^{(1) \,DY}_{q, NS} = C_F \left( \frac{\pi^2}{2} - \frac{7}{2} \right) \;\;.
\end{equation}
Inserting Eq.~(\ref{h1dyns}) in Eq.~(\ref{b2dep}), and using the DY coefficient
in Eq.~(\ref{b2dy}), we then obtain the corresponding coefficient of the
quark form factor:
\beeq
\!\! B_{q,NS}^{(2)} &\!\!\!\!=\!\!\!\!& B_q^{(2) \,DY} 
- \beta_0 H^{(1) \,DY}_{q, NS} \nn \\
\label{b2ns}
&\!\!\!\!=\!\!\!\!&C_F^2\left(\frac{\pi^2}{4}-\frac{3}{16}-3\zeta_3\right)
+C_F\,C_A\left( - \frac{11}{72}\pi^2 - \frac{13}{16} +\frac{3}{2}\zeta_3\right)
+ C_F\,N_f\left( \frac{1}{8} + \frac{\pi^2}{36}\right) .
\eeeq
The reason why this scheme can be considered `more physical' is that the first
moments of the 
NS cross sections are collinear-safe and, hence, free from ambiguities related
to the regularization procedure of collinear singularities. As is known
\cite{Kodaira:1982nh, Catani:1988vd, Catani:1991rr}, the process-independent
form factor coefficients 
in Eq.~(\ref{ffcoef}) measure the physical intensity of soft ($A_c^{(1)}$ and 
$A_c^{(2)}$) and collinear $(B_c^{(1)})$ radiation. Since the 
coefficient $B_{q,NS}^{(2)}$ in Eq.~(\ref{b2ns}) appears in the form factor
of an infrared {\em and} collinear safe quantity, it should measure the
intensity of the collinear radiation from quarks at the second order in $\as$
and it should enter in the form factor of other infrared and collinear safe 
quantities. For example, an analogous transverse-momentum form factor  
controls the back-to-back region of the energy--energy correlation (EEC)
in $e^+e^-$ annihilation \cite{Dokshitzer:1999sh}. The coefficient in 
Eq.~(\ref{b2ns}) should directly be related to the corresponding 
coefficient $B^{(2)}$ of the EEC, modulo (possible) effects due to the 
difference between the space-like kinematics of the transverse-momentum
cross sections and the time-like kinematics of $e^+e^-$ annihilation.

The generalization of the `NS resummation scheme' to $gg$-fusion processes 
can be obtained, for instance, by fixing the overall normalization of the 
gluon coefficient function $C_{gg}(\as,z)$ in Eq.~(\ref{unw}) 
in such a way that its first moment vanishes. Note, however, that in the gluon
(or, more generally, flavour-singlet) channel there is no anologue of the
fermion-number conservation rule. Flavour-singlet transverse-momentum
cross sections are collinear unsafe quantities. Thus, a `physical' 
(collinear-safe) interpretation of the higher-order coefficients of the function
$B_g(\as)$ of the gluon form factor is less straightforward than in the quark
channel.

The coefficient $B_c^{(1)}$ in Eq.~(\ref{ffcoef}) coincides with the 
coefficient of the end-point contribution 
(the term proportional to $\delta(1-z)$) to the LO Altarelli--Parisi
probability $P_{cc}^{(1)}(z)$. As can be noticed from Eqs.~(22) and (23) in
Ref.~\cite{defgraz}, the coefficient
$B_{q,NS}^{(2)}$ in Eq.~(\ref{b2ns}) (and the coefficients 
$B_q^{(2) \,DY}$ and $B_g^{(2) \,H}$ in Eqs.~(\ref{b2dy}) and (\ref{b2h}))
does not coincide with that of the
end-point contribution to the NLO Altarelli--Parisi probability
$P_{cc}^{(2) \,\msbar}(z)$
in the \ms\ factorization scheme. This should not 
be surprising, because beyond LO 
the Altarelli--Parisi probabilities are not collinear-safe 
quantities: they depend on the regularization procedure
of the collinear singularities and on the factorization scheme. This
interpretation is confirmed by the fact that the various coefficients
$B_c^{(2)}$ differ from one another by terms proportional to the first 
coefficient $\beta_0$ of the $\beta$-function, as should be expected 
from factorization-scheme dependence.
Using Eqs.~(\ref{cdep})--(\ref{b3dep}),
it is obviously possible to introduce an `\ms\ resummation scheme' such that
the function $B_c(\as)$ of the process-independent form factor
in Eqs.~(\ref{formfact}) and (\ref{unw}) coincides by definition
with the perturbative function $B_{c, \msbar}(\as)$ that controls the 
end-point contribution to the all-order Altarelli--Parisi probabilities 
in the \ms\ factorization scheme. 

Note that our discussion on the relation between Eqs.~(\ref{nonunw}) and 
(\ref{unw}) and on the resummation-scheme ambiguity does not imply that the 
two equations are equivalent.
In fact, the process-independent resummation framework contains
more information than its customary process-dependent version.
To evaluate resummed cross section within the framework of Eq.~(\ref{nonunw}),
we should compute two functions ($S_c^F$ and $C_{ab}^F$) for each process $F$.
Once a `resummation scheme' has been defined, a similar evaluation  
by using Eq.~(\ref{unw}) requires the computation of two universal
functions ($S_c$ and $C_{ab}$) and of a single additional function $(H_c^F)$
for each process. 

Of course, we can still continue to use Eq.~(\ref{nonunw}). In this case, 
we can exploit the additional information contained in Eq.~(\ref{unw}) by using
Eqs.~(\ref{cff}) and (\ref{bff}) (or their perturbative expansions in 
Eqs.~(\ref{c1ff})--(\ref{b3ff})) to relate form factors and coefficient functions
for different processes.

Independently of the resummation formula that is actually used, once the
form factor coefficient $B_c^{(2)}$ has been computed for a single
process by performing a calculation at relative order $\as^2$, no further
${\cal O}(\as^2)$-calculation is necessary for its universal
implementation in all processes. The implementation can simply be performed
by computing the coefficient functions at ${\cal O}(\as)$ and then using
Eqs.~(\ref{c1dep}) and (\ref{b2dep}) (or Eqs.~(\ref{c1ff}) and (\ref{b2ff})). 
Analogously, using Eqs.~(\ref{c2dep}) and (\ref{b3dep}) (or Eqs.~(\ref{c2ff}) 
and (\ref{b3ff})), the universal implementation of the form factor 
coefficient $B_c^{(3)}$ requires its ${\cal O}(\as^3)$ computation in a single
process and the evaluation of the coefficient functions at ${\cal O}(\as^2)$
for the various processes. This procedure extends to any higher orders.

At present, owing to the results in Eqs.~(\ref{b2dy}) and (\ref{b2h}), 
the NNLL coefficient $B_c^{(2)}$ can be included in all calculations
of transverse-momentum cross sections in hadron collisions. Its inclusion
cannot be regarded as the extension of these resummed calculations
to full NNLL accuracy, because the corresponding coefficient $A_c^{(3)}$
is not yet known. Nonetheless, the knowledge of $B_c^{(2)}$ can certainly be 
used to improve the matching with fixed-order calculations at high $\qt$.
The matching can be performed by supplementing the resummed cross section
with the finite contribution on the right-hand side of Eq.~(\ref{Fdec}).
The finite contribution is obtained, from the complete
calculation at a fixed order in $\as$, by subtraction of the corresponding
perturbative terms already included in the resummed component. If $B_c^{(2)}$
is included in this component, the remaining finite component
at relative order $\as^2$ contains no $\ln Q^2/\qt^2$ terms. Thus, the latter
can be inserted in Eq.~(\ref{Fdec}) uniformly with respect to $\qt$,
that is without including logarithmic terms that spoil the 
convergence of the matched result in the small-$\qt$ region.

In this paper we have not discussed how non-perturbative effects
(see Refs.~\cite{Ellis:1997sc}--\cite{Kulesza:1999gm}
and references therein) affect the transverse-momentum cross sections.
Although the process-independent resummation formula (\ref{unw}) can
be recast in the form of Eq.~(\ref{nonunw}) at the perturbative level,
it is not evident whether the two formulations are equally suitable 
to deal with non-perturbative contributions. In particular,
since the form factor $S_c^F$ and the coefficient functions 
$C_{ab}^F$ in Eq.~(\ref{nonunw}) are 
process-dependent\footnote{We remind the reader that the
process dependence can kinematically be quite non-trivial when the 
transverse-momentum cross section in not fully
integrated over the kinematical variables of the final-state system $F$.},
a simple and uniform implementation of process-independent
non-perturbative effects (such as those due to the initial intrinsic $k_t$ of the
partons in the colliding hadrons) in Eq.~(\ref{nonunw}) can be very involved.
The process-independent version (\ref{unw}) of the resummed cross section
can help to consistently introduce and implement non-perturbative effects
in transverse-momentum distributions.

We conclude by discussing the dependence of the resummation formulae on
the renormalization and factorization scales $\mu_R$ and $\mu_F$.
This dependence is often parametrized by introducing some arbitrary
coefficients, $C_1, C_2, C_3,$ as suggested in 
Refs.~\cite{Collins:1981uk, Collins:1985kg}. This is a perfectly sensible
and reasonable procedure to try to estimate the effect of higher-order
corrections, but it does not exactly correspond to the procedure that is
usually followed in fixed-order perturbative calculations. In the case of
soft-gluon resummed calculations for event shapes and jet rates in $e^+e^-$ 
annihilation and for threshold contributions in hadronic cross sections, a
different procedure was introduced in Refs.~\cite{Catani:1991kz} and 
\cite{Bonciani:1998vc}. The latter procedure closely matches the use of
$\mu_R$ and $\mu_F$ in fixed-order calculations, and it can directly be applied
(see, e.g., Refs.~\cite{Catani:1991kz, Bonciani:1998vc} 
and several other references quoted in Section~5 of Ref.~\cite{Catani:2000jh})
to compare the scale dependence of resummed and fixed-order
predictions. Here, we would like to note that the procedure of 
Refs.~\cite{Catani:1991kz, Bonciani:1998vc} can be introduced also in the case
of the transverse-momentum distributions\footnote{A similar observation can be found in Ref.~\cite{Frixione:1999dw}.}.

To this purpose, we first perform the replacement
$f_{a/h}(x,b_0^2/b^2) \to f_{a/h}(x,\mu_F^2)$ in Eq.~(\ref{resgen}) 
and we rewrite the resummed component of the transverse-momentum
cross section as
\beeq
\label{resgenmu}
\left[ \frac{d\sigma_{F}}{dQ^2 \;d\qt^2} \right]_{\res} &=& \sum_{a,b}
\int_0^1 dx_1 \int_0^1 dx_2 \;f_{a/h_1}(x_1,\mu_F^2) 
\; f_{b/h_2}(x_2,\mu_F^2) \nn \\
&\cdot& \int_0^\infty db \; \frac{b}{2}\; J_0(b \qt) 
\; W_{ab}^{F}(x_1 x_2 s; Q, b, \mu_F) \;\;.
\eeeq
To obtain Eq.~(\ref{resgenmu}) we have simply used the scale dependence of
the parton densities as given by
\begin{equation}
f_{a/h}(x,b_0^2/b^2) = \sum_b \int_x^1 \frac{dz}{z}
\;U_{ab}(z;b_0^2/b^2,\mu_F^2)  \;f_{b/h}(x/z,\mu_F^2) \;\;,
\end{equation}
where $U_{ab}(z;b_0^2/b^2,\mu_F^2)$ is the customary evolution operator
matrix obtained by solving the Altarelli-Parisi evolution equations to the 
required logarithmic accuracy. The resummed partonic cross section in 
Eq.~(\ref{resgenmu}) is thus obtained from that in Eq.~(\ref{unw}) (or 
(\ref{nonunw})) by replacing the coefficient functions
$C_{ca}(\as(b_0^2/b^2), x)$ (or $C_{ca}^{F}(\as(b_0^2/b^2), x)$) with the
following convolution
\begin{equation}
C_{ca}(\as(b_0^2/b^2), x; \mu_F^2) = \sum_b \int_x^1 \frac{dz}{z} \;
C_{cb}(\as(b_0^2/b^2), z)
\;U_{ba}(z;b_0^2/b^2,\mu_F^2) \;\;.
\end{equation}

Note that, unlike the parton densities
in Eq.~(\ref{resgen}), those in Eq.~(\ref{resgenmu}) do not
depend on the impact parameter $b$, but only on the factorization scale 
$\mu_F$. Expression (\ref{resgenmu}) can thus be useful in practical
calculations, because it avoids the numerical integration of the
parton densities with respect to the impact parameter. The only factor
of the resummed cross section that has to be integrated over $b$ is now
the perturbative component $W_{ab}^{F}(x_1 x_2 s; Q, b, \mu_F)$.

Then, to proceed as in Refs.~\cite{Catani:1991kz, Bonciani:1998vc}, we
simply observe that we can choose $\mu_F$ and $\mu_R$ to be of the order of 
the hard scale $Q$. Thus, $W_{ab}^{F}(x_1 x_2 s; Q, b, \mu_F)$ contains only
an additional large parameter, $L\equiv \ln Q^2b^2$, and it can be written as
\beeq
\label{wexp}
W^{F}(Q, b, \mu_F) &\sim& \exp \left\{ L \;f_1\left(\as L \frac{}{}\right) 
+ f_2\left(\as L; \frac{Q}{\mu_R}, \frac{Q}{\mu_F}\right) + 
\as f_3 \left(\as L;\frac{Q}{\mu_R}, \frac{Q}{\mu_F} \right)+ \dots \right\} 
\nn \\
&\cdot&  \sigma^F(Q^2, \as; Q^2/\mu_R^2) \;\;,
\eeeq
where $\as = \as(\mu_R^2)$. 
The notation on the right-hand side is symbolical, because
we have understood the dependence on the momentum fractions~\footnote{The
convolutions with respect to the momentum fractions can as usual be diagonalized
by taking $N$-moments and working in Mellin space.} 
$z_1,z_2$ and on 
the parton indices $a,b=q,{\bar q},g$. In particular, the exponential should
be understood as an exponential matrix that depends on the parton indices.

The form in Eq.~(\ref{wexp}) is straightforwardly obtained by using 
the exponentiated
result in Eq.~(\ref{formfact}) for the form factor and the customary 
exponential form for the evolution operator $U_{ab}(z;b_0^2/b^2,\mu_F^2)$.
In the exponent, the function $Lf_1$ resums all the LL contributions 
$\as^n L^{n+1}$, the function $f_2$ resums the NLL terms $\as^n L^{n}$,
while $\as f_3$ contains the NNLL terms $\as^n L^{n-1}$, and so forth.
The functions $f_i$ are normalized as $f_i(\as L=0) = 0$
and can easily be obtained in analytic form. 

Although it is symbolical,
the form of Eq.~(\ref{wexp}) is sufficient to make our point. Indeed, it is
exactly in the same form as the expressions that are obtained by
performing soft-gluon resummation for $e^+e^-$ event shapes and hadronic cross
sections near threshold. Recasting the resummed transverse-momentum
distributions in the form of Eq.~(\ref{wexp}), 
the theoretical accuracy of the resummed calculation can be investigated as 
in fixed-order calculations, by varying $\mu_F$ and $\mu_R$
around the value $Q$ of the typical hard scale. 
Future studies along these lines (for instance, comparisons of the scale 
dependence of resummed and fixed-order calculations) could be useful 
to increase our confidence in the quantitative reliability of resummed
predictions for transverse-momentum cross sections.

\end{document}